# SELF-ASSEMBLY AND DYNAMICS OF MAGNETIC HOLES

A.T. SKJELTORP[1,2], J. AKSELVOLL[1,2],
K.de LANGE KRISTIANSEN[1,2], G. HELGESEN[1], R. TOUSSAINT[2,3],
E.G. FLEKKØY[2], and J. CERNAK[4]
[1]*Institute for Energy Technology, POB 40, NO-2027 Kjeller, Norway*
[2]*Department of Physics, University of Oslo, NO-0316 Oslo, Norway*
[3]*Department of Physics, NTNU, NO-7491 Trondheim, Norway*
[4]*University P.J. Safarik Kosice, Biophysics Department, Jesenna 5, SK-043 54 Kosice, Slovak Republic*

**Abstract**

Nonmagnetic particles in magnetized ferrofluids have been denoted magnetic holes and are in many ways ideal model systems to test various forms of particle self assembly and dynamics. Some case studies to be reviewed here include:
- Chaining of magnetic holes
- Braid theory and Zipf relation used in dynamics of magnetic microparticles
- Interactions of magnetic holes in ferrofluid layers

The objectives of these works have been to find simple characterizations of complex behavior of particles with dipolar interactions.

## 1. Ferrofluids

Figure 1 shows the characteristic features of ideal ferrofluids on different length scales. Ferrofluids consist of coated magnetic nanoparticles dispersed in a carrier liquid. The nanoparticles are so small that they contain only one magnetic domain, i.e. at this length scale it is not energetically favorable to break up into domains as in ordinary bulk magnets. Ferrofluids are in fact an early success story in the commercialization of nanotechnology [1]. In the 1970s, ferrofluids were adopted by the disk drive industry as near-zero friction bearings. Today, ferrofluid bearings are a key component in greatly reducing the incidence of hard-disk failure, and there is also a wide range of other ferrofluid applications. Ferrofluids are often denoted magneto-rheological fluids (MRF) as they may exhibit fast, strong and reversible changes in their rheological properties when a magnetic field is applied. MRF are similar to electro-rheological fluids, but normally much stronger, and more stable and easier to use. Applications of MRF include clutches, damping systems in passenger vehicles, brakes, controllable friction damper that decreases the noise and vibration in washing machines, and seismic mitigation MRF damping systems protecting buildings and bridges from earthquakes and windstorms [2]. Ferrofluids exhibit also many exotic macroscopic phenomena as exemplified in Figure 2.

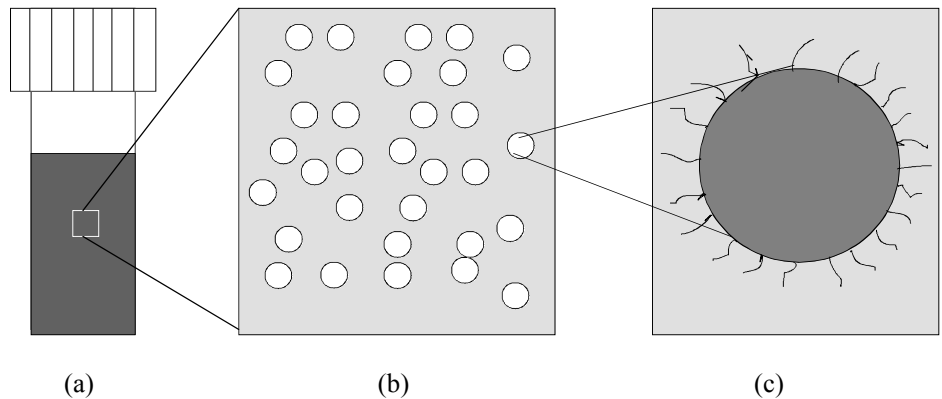

*Figure 1.* Schematic picture of ferrofluid on three length scales:
a: On a macroscopic length scale, it resembles an ordinary liquid with a uniform magnetization.
b: On a colloidal length scale, solid nanoparticles dispersed in a liquid.
c: Each particle consists of a single domain magnetic core, e.g. iron oxide, and a surface grafted with polymer chains, particle size ~ 10 nm.

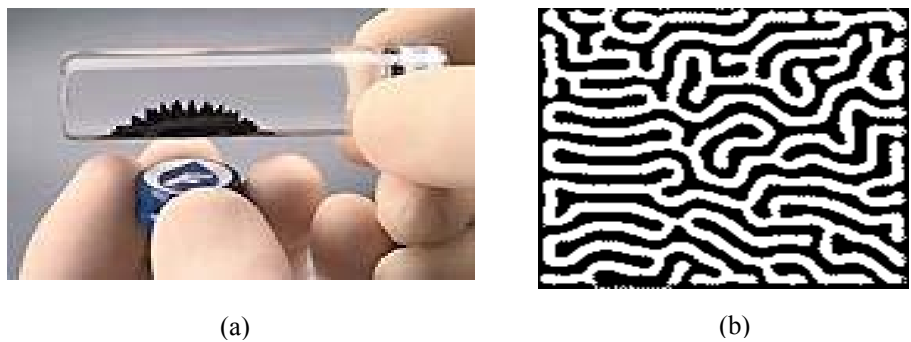

*Figure 2.* (a) Surface instability of ferrofluid subject to an external magnetic field (http://www.ferrotec.com/) and (b) a ferrofluid meander in a thin layer.

## 2. Magnetic Holes

Monodisperse polystyrene spheres dispersed in ferrofluid provide a convenient model system for the study of various order-disorder phenomena [3]. The basis for this is that the spheres displace ferrofluid and behave as magnetic holes [4] with effective moments equal to the total moment of the displaced ferrofluid. The spheres are much larger (1-100 μm) than the magnetic particles (typically 0.01 μm) in the ferrofluid and the spheres therefore move around in an approximately uniform magnetic background. By confining the spheres and ferrofluid between closely spaced microscope slides an essentially two-dimensional many-body system of interacting particles is obtained. This offers the possibilities of observing directly through a microscope a wide range of nonlinear dynamic phenomena and collective processes, as they are easy to produce and to manipulate with external magnetic fields. A simplifying feature of magnetic holes is that their magnetic moments are collinear with an external field at any field strength. This is in contrast with magnetic particles dispersed in non-magnetic fluids where random orientation of the magnetic moments complicates the theoretical treatment of the dynamic and static properties of the particles.

The basic principle for magnetic holes is shown in Figure 3. It is in some sense a magnetic analogue of Archimedes' principle. When a non-magnetic particle is dispersed in a magnetized ferrofluid ($H > 0$), the void produced by the particle possesses an effective magnetic moment $M_V$ equal in size but opposite in direction to the magnetic moment of the displaced fluid

$$M_V = -V\chi_{eff}H, \tag{1}$$

where $V$ is the volume of the sphere and $\chi_{eff} = 3\chi/(3+2\chi)$ is the effective volume susceptibility of the ferrofluid. The interaction energy between two spheres with a centre-to-centre separation $d$ is given approximately by the dipolar interaction

$$U = \frac{M_v^2(1-3\cos^2\theta)}{d^3}. \tag{2}$$

Here, θ is the angle between the line connecting the centres of the spheres and the direction of the field. Figure 3 illustrates that if the centers of two holes are collinear, they will attract each other, while two holes side by side will repel. Typical examples of structures formed for the two field orientations are shown in Figure 4.

A detailed description of the interaction between the spheres in a lattice is quite complicated. Since the dipolar interaction is of relatively long range, the direct particle interaction goes far beyond the nearest neighbours. In addition, there is an indirect particle-particle interaction mediated via the glass plates confining the system. This "image dipole" effect is caused by the change in the magnetic permeability across the glass plates. Thus the spheres also interact with their image dipoles situated at the opposite side of the ferrofluid-glass interface. This effect causes the lattice to be situated precisely midway between the upper and lower glass plate. As the plate separation is typically 50-100% larger than the diameter of the spheres, the dipole image contribution is relatively small (typically 10% or less) compared with the interaction energy between the real dipoles. Even in the presence of the dipole images, this experimental system may still be considered to be two-dimensional.

It is possible to obtain a thermodynamic system by using small spheres (diameter less than 2 μm) since Brownian motion introduces fluctuations into the system. The controlling parameter for the stability of the structure formation is the ratio of the dipolar energy to the thermal energy:

$$\Gamma = \frac{M_v^2}{d^3 k_B T} \tag{3}$$

where $d$ is the (centre-to-centre) separation of the spheres and $k_B$ is Boltzmann's constant.

The ability to design and modify the effective interactions in this system enables studies of various phenomena discussed in the next sections.

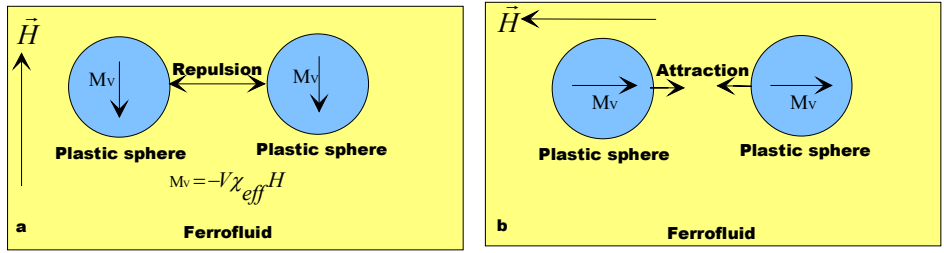

*Figure 3.* The principle of a magnetic hole in a soft magnet as discussed in the text. (a) Two holes side by side will repel each other. (b) Two holes with the centers collinear with the field lines will attract each other.

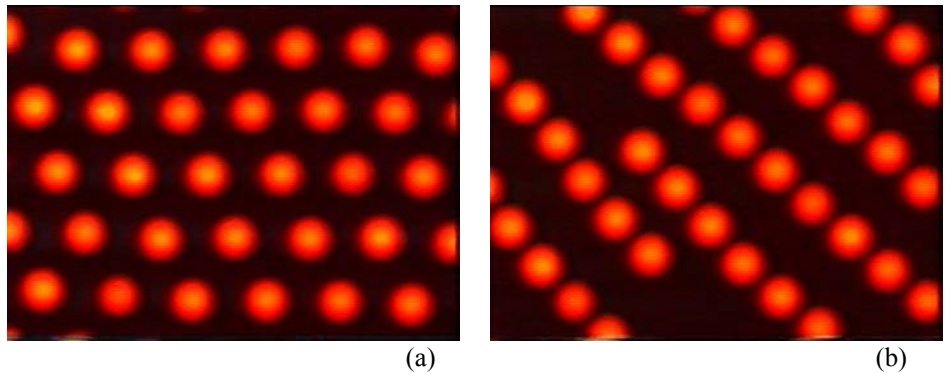

*Figure 4.* Structures of 10 μm diameter spheres formed by a magnetic field (a) normal to the layer and (b) parallel to the layer.

## 3. Aggregation and Chaining of Magnetic Holes

Experimental studies of field induced colloidal aggregation have been carried out earlier, e.g., with paramagnetic microspheres [5,6], magnetic nano-particles [7,8], and electric field-induced association of dielectric particles [9]. The results have essentially confirmed the scaling behavior of the mean cluster size as a function of time [10-12] and it has been possible to scale the temporal size distribution of clusters [5,8,13] into a single universal curve as predicted by dynamic scaling theory [14].

In the present work we have studied the chain formation of non-magnetic microspheres [15] dispersed in thin layer of ferrofluid [16] induced by external magnetic fields, see Figure 5.

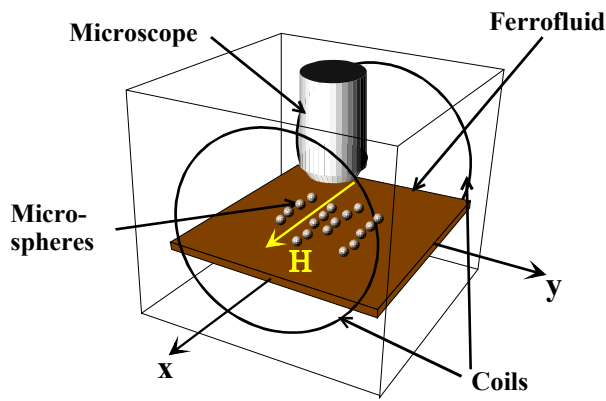

*Figure 5.* Schematic experimental set-up.

The main advantage of this system is the possibility to tune the strength of the particle-particle interactions. In our case we can create experimental conditions that are close to an ideal point-dipole system, i.e, the particles are spherical, monodisperse, and their resulting induced magnetic moments are oriented in the direction of the external magnetic field. In order to study the importance of dipole-dipole interaction and Brownian motion relative to non-Brownian ballistic drift, we used microspheres with different diameters $d$ = 1.9, 4.0, and 14 µm. In zero external fields the diffusive Brownian motion of the 1.9 µm spheres is clearly visible in the microscope. However, the diffusion of the 14 µm spheres can only be seen by comparing images taken at typically 30 s time lap. The experiments were done with low volume fractions of microspheres, corresponding to coverage of a few percent. Magnetic fields in the range $H$ = 4 – 16 Oe were used. The combination of these fields with the three particle sizes gave rise to values of $\Gamma$ in the range 8 - $10^4$. A typical example of the aggregates that were formed is shown in Figure 6.

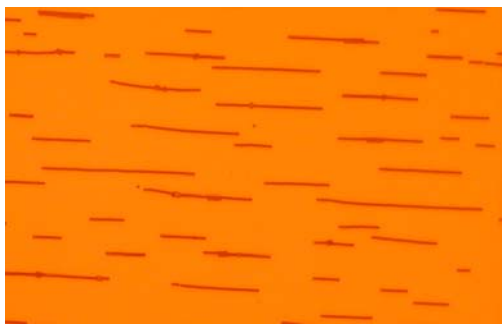

*Figure 6.* An example of the straight aggregates formed by 4 µm spheres in 10 Oe magnetic field after about 2 hours.

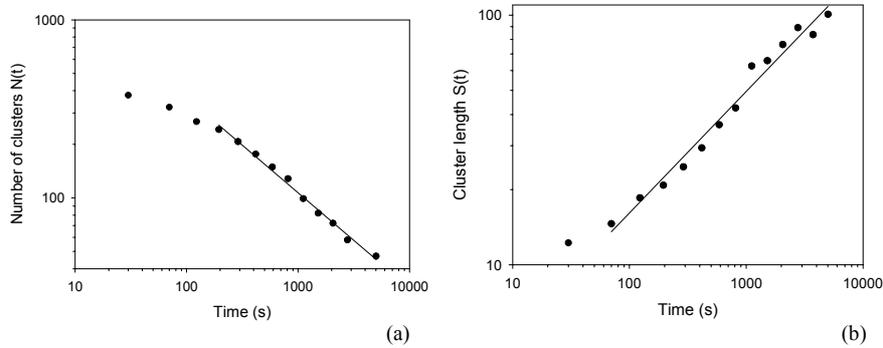

*Figure 7.* (a) The number of clusters and (b) cluster length versus time for aggregation of 10 μm spheres in a 10 Oe magnetic field. The solid lines are fits to a power law with exponents (a) $z' = 0.53$ and (b) $z = 0.50$.

In order to characterize the aggregation process in more detail, the length $s$ of any cluster was determined at different time intervals $t$ relative to the initial time $t = 0$ when the field was turned on. The time dependences of the number of clusters $N(t)$ and mean cluster size (length) $S(t)$ for a typical experiment with $d = 4$ μm spheres are shown in Figure 7. We see that the data asymptotically follow the power laws $N(t) \sim t^{-z'}$ and $S(t) \sim t^z$ with scaling exponents $z' = 0.53$ and $z = 0.50$ for $t > 100$ s. Using the dynamical scaling relations for cluster-cluster aggregation [12,14], it was possible to collapse all the cluster size distributions $n_s(t)$ to a scaling function curve [17].

## 4. Braid Theory and Zipf Relation used in Dynamics of Magnetic Microparticles

Cooperative behavior of interacting microparticles is of great interest both from a fundamental and a practical point of view. The rank-ordering statistics is one way of analyzing such collective diffusive systems [18]. Here we report some results from our studies of a colloidal system of magnetic holes with an experimental set-up like that in Figure 5. Intricate motions of particles in a plane can be illustrated as entangled lines in three-dimensional space-time, Figure 8, and braid theory gives a compact description of these lines [19,20].

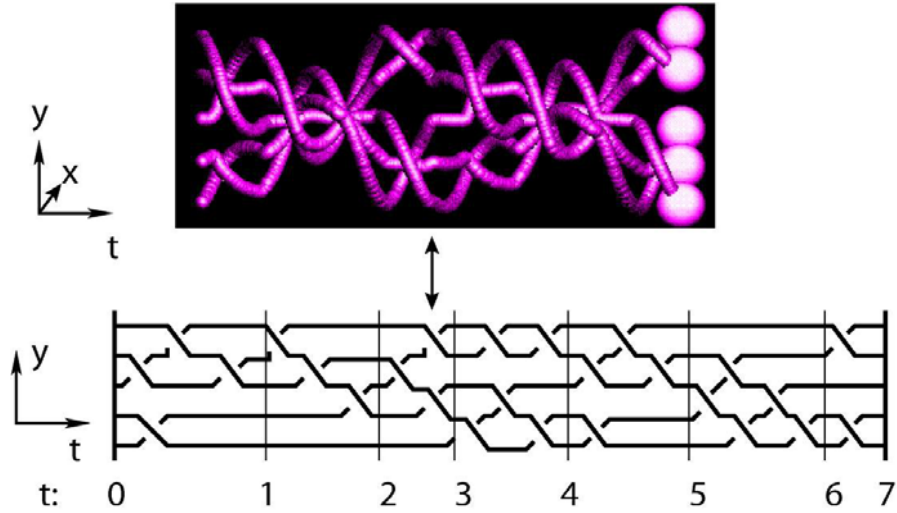

*Figure 8.* The two-dimensional motion of 5 microspheres is extended to three-dimensional space-time with the trace of each particle in time (upper). Projecting onto (y,t)-axis gives the braid for the motion (lower).

The magnetic holes system consists of 50 μm diameter diamagnetic, polystyrene micro-spheres [15] immersed in a thin layer of ferrofluid [16]. In an external planar, elliptically polarized rotating magnetic field *H* a microsphere introduces a magnetic hole with apparent magnetic moment given by Eq. (1). The motions of the microspheres can easily be simulated [7] by considering, to first order, the magnetic dipole-dipole interaction between each pair of microspheres and the viscous force on each microsphere.

By varying the angular velocity of the magnetic field and its anisotropy, we observe a rich diversity of patterns. Here we will focus on non-ordered behavior. The structure and complexity of the braids formed by the traces of the moving particles can be captured by calculating a few characteristic numbers. One of these is found by transforming the braid into a positive permutation braid [22] and use that as a measure of the systems dynamical mode. Then we apply rank-ordering statistics on these permutation braids, i.e., count all the different modes and rank them. Figure 9 shows a plot of the frequency of occurrence of the permutation braid *ϕ(r)* vs. rank *r*. This graph show a power law dependence, according to the Zipf relation [23] $\phi(r) \sim r^{\gamma}$ with $\gamma = 1.2$ for about two decades in *r*. Attempts to understand the origin of this relation were connected to the structure of hierarchical systems [24].

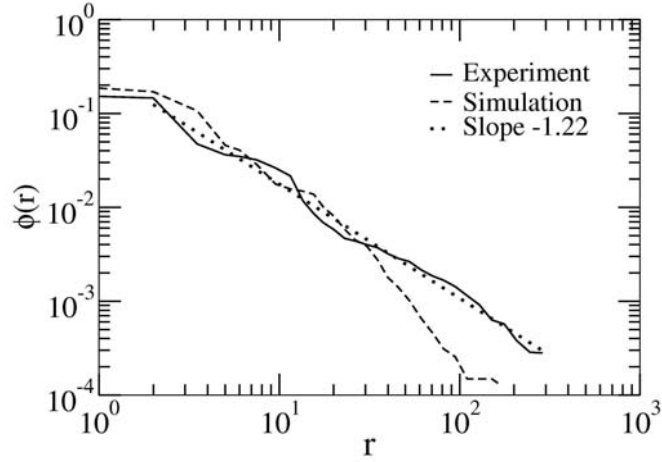

*Figure 9.* The relative occurrences of braid permutations $\Phi(r)$ against its rank $r$ found in the dynamics of 7 microspheres.

Another characteristic number is the writhe $Wr(t)$. It is equal to the number of over-crossings minus the number of under-crossings, starting with value zero at time zero. Here the time $t$ is measured in units of half periods of the external rotating magnetic field. One measure of the fluctuation in the braid pattern is the variation of the writhe around its mean:

$$\delta Wr(t,\tau) = Wr(\tau + t) - Wr(\tau) - t\langle \delta Wr(t)\rangle, \qquad (4)$$

where $\langle \delta Wr(t)\rangle = \frac{1}{N}\sum_{t=1}^{N}[Wr(t) - Wr(t-1)]$. The variance $\sigma^2(t)$ is then:

$$\sigma^2(t) = \langle [\delta Wr(t,\tau) - \langle \delta Wr(t,\tau)\rangle]^2\rangle. \qquad (5)$$

For diffusive processes: $\sigma^2(t) \sim t^\beta$, with $\beta = 1$ for ordinary diffusion. When $\beta$ is smaller or larger than 1, the diffusion is anomalous and is denoted as subdiffusion and superdiffusion, respectively.

We have found that the variance of the writhe fluctuations in experiments behave as a power law as shown in Figure 10, with $\beta = 1.66$, clearly indicating anomalous diffusion. Simulations based on the simplest model with magnetic and viscous forces showed good agreement with $\beta = 1.82$. Theoretical work using generalized Langevin and Fokker-Planck equations has predicted anomalous diffusion with $\beta = 1.66$ [25], in good agreement with our results.

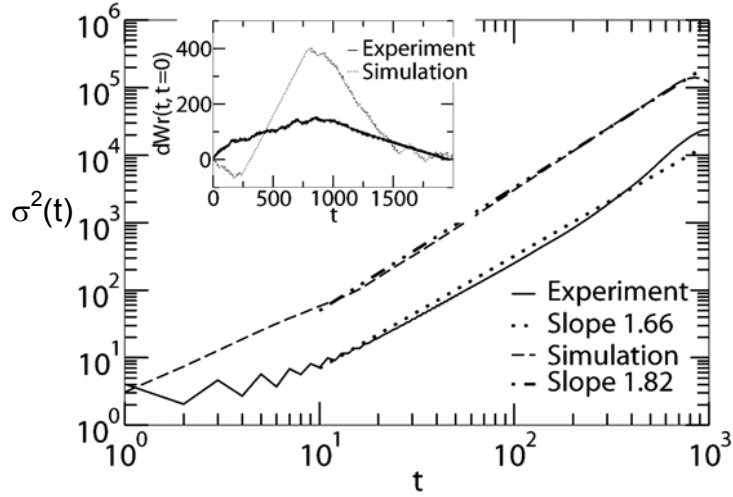

*Figure 10.* The variances $\sigma^2(t)$ of the writhe fluctuations (Eq. 5) for the dynamics of 7 microspheres. The inset illustrates how the writhe of the system evolves with time.

## 5. Interactions of Magnetic Holes in Ferrofluid Layers

So far there has been no satisfactory theoretical description of the effective interactions between magnetic holes in rotating magnetic fields composed of a high frequency rotating inplane component and a constant normal one, and the existence of the observed stable configurations of holes with a finite separation distance [26] has remained unexplained.

Focusing on the boundary conditions of the magnetic fields along the confining plates, we have derived analytically the pair interaction potential in such oscillating fields, and demonstrated for a wide range of conditions the existence of a secondary minimum whose position depends continuously on the ratio $\beta$ between the out of plane $H_\perp$ and inplane $H_\parallel$ field components [27]. We compare this theory with experiments where the motion of a pair of particles (holes) is followed.

In presence of a far-range field $\vec{H}$ in a ferrofluid of susceptibility $\chi$, each hole generates a dipolar perturbation of dipolar moment equal to the opposite of the displaced ferrofluid, $\vec{\sigma} = -V\chi_{eff}\vec{H}$, where $V$ is the volume of the hole, and $\chi_{eff} = 3\chi/(3+2\chi)$ is an effective susceptibility including a demagnetization factor rendering for the boundary conditions of the magnetic field along the spherical particle boundary [4,28]. The susceptibility contrast between the ferrofluid and the two plane nonmagnetic confining plates leads to a different dipolar field perturbation than the infinite medium expression. According to the image method [29], the boundary conditions for the magnetic field along the plates are fulfilled with an addition of an infinite series of dipole images to the infinite space expression of the dipolar field emitted in an unbounded medium. The images are constructed as mirror images in the plane boundaries of the initial dipoles or of some previous image, multiplying the magnitude of the dipole at each mirror symmetry operation by an attenuation factor $\kappa = \chi/(\chi + 2)$ - see Figure 11.

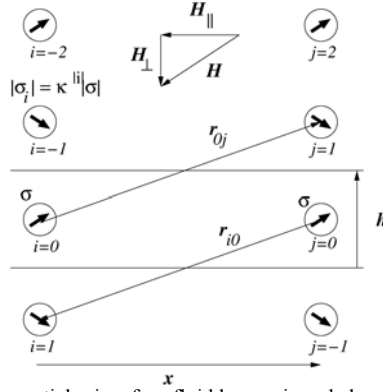

*Figure 11.* Pair of non magnetic particles in a ferrofluid layer, viewed along the confinement plates, and the series of dipole images accounting for the boundary conditions of the magnetic field along the plates.

The instantaneous interaction potential between a pair of confined holes can then, similarly to Eq. (2), be expressed as

$$U = \frac{\mu}{8\pi} \sum_{i \neq j} \left[ \frac{\vec{\sigma}_i \cdot \vec{\sigma}_j}{r_{ij}^3} - 3 \frac{(\vec{\sigma}_i \cdot \vec{r}_{ij})(\vec{\sigma}_j \cdot \vec{r}_{ij})}{r_{ij}^5} \right], \qquad (6)$$

where $\mu$ is the ferrofluid's permability. The *i*-index runs over both the source and image dipoles, and the *j*-index runs only over the two source dipoles. A detailed analysis of the above shows that the dominant effect for the force components normal to the plates, is the interaction between a particle and its own mirror images, which stabilizes the particles midway between the plates.

Decomposing the instantaneous field in its inplane and normal components $\vec{H}_\perp$ and $\vec{H}_\parallel$, we define the ratio of their magnitudes $\beta = H_\perp / H_\parallel$, the particle diameter and interplate separation respectively as $a$ and $h$, and the scaled separation as $x = r/h$. At the field rotation frequencies used here, $f = 10 - 100$ Hz, which exceeds the inverse viscous relaxation time, the motion of the holes can be neglected during a field rotation, and an effective, time-averaged interaction potential can be obtained by averaging over a full rotation of the magnetic field, while the slowly varying separation vector is maintained constant. This leads to the dimensionless effective interaction potential

$$u(x) \equiv \frac{144 h^3 \overline{U}}{\mu \pi a^6 \chi_e^2 H_\parallel^2} = \sum_{l=-\infty}^{+\infty} \kappa^{|l|} \left[ \frac{1+(-1)^{|l|}\beta^2}{(x^2+l^2)^{3/2}} - 3\frac{(-1)^{|l|}l^2\beta^2 + y^2/2}{(x^2+l^2)^{5/2}} \right]. \quad (7)$$

The term $l=0$ corresponds to the source-source interaction term, already used in previous studies [4,26], the others to interactions between a particle and the images of the other one. For all existing ferrofluids, $\kappa$ is sufficiently smaller than unity so that the prefactor $\kappa^{|l|}$ ensures that the three first images are enough to get a relative precision better than 1% for the potential and its derivatives.

For the micrometer sized particles used here, inertial terms can be neglected, and a characteristic viscous relaxation time can be obtained as the time to separate two particles initially in contact by a distance equal to their size. Balancing a Stokes drag with the magnetic interaction forces derived from the potential above and retaining the main term in Eq.(7), leads to the estimate $T_c = 144\eta / \mu \chi_e^2 H^2 \approx 5s$, for the ferrofluid ($\eta = 9 \cdot 10^{-3}$ Pa.s, $\mu = \mu_0(1+\chi)$, $\chi = 1.9$, $\mu_0 = 4\pi \cdot 10^{-7}$ H.m$^{-1}$) and a typical field $H = 10$ Oe.

For a given ferrofluid and field, this central potential can be of four possible types as illustrated in Figure 12. At low normal field $\beta < \beta_m$, the interactions are purely attractive up to contact; at higher ones $\beta_m < \beta < \beta_c$, a secondary minimum at finite distance appears, later on in a regime $\beta_c < \beta < \beta_u$ this minimum becomes the only one, and ultimately interactions are purely repulsive for $\beta_u < \beta$.

From Eq. (7), the separating values of $\beta$ can be shown to be $\beta_c = 1/\sqrt{2}$, and $\beta_u = \beta_c (1+\kappa)/(1-\kappa)$ which is a growing function of the susceptibility. $\beta_m$ is a function of the susceptibility which decreases regularly from $\beta_c$ at zero susceptibility to 0 at an infinite one. For the ferrofluid used in the present experiments $\beta_m \approx 0.55$ and $\beta_u \approx 2.05$. Neglecting the susceptibility contrast along the plates, i.e., terms with $l \neq 0$ in Eq. (7), would correspond to the limiting case $\kappa = 0$, where these three separating values merge, and the interactions are either purely attractive or repulsive. The presence of this susceptibility contrast is thus essential to trap the particles at a given equilibrium distance in this type of field.

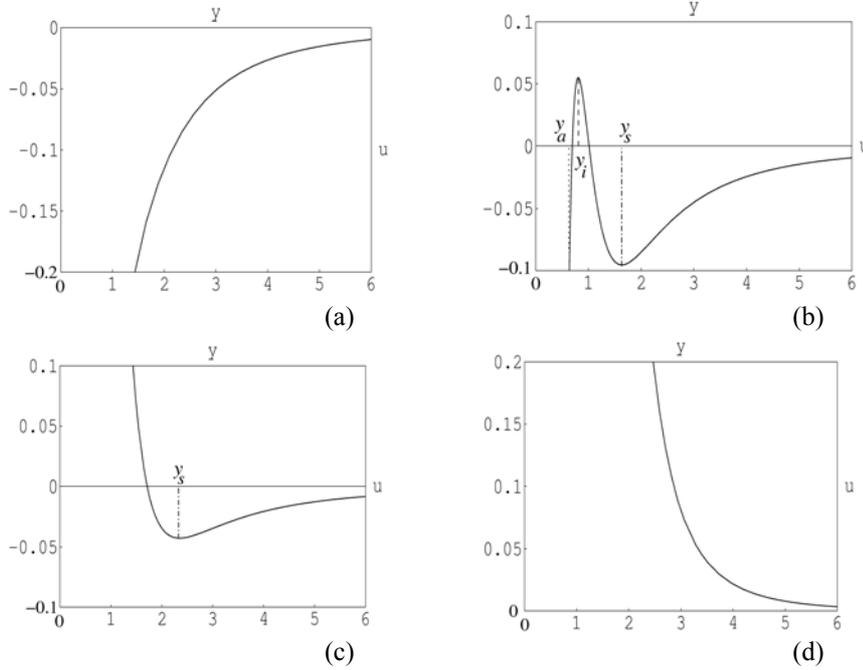

*Figure 12*. Four possible types of the interaction potential: (a) β<β$_m$ purely attractive interactions, (b) β$_m$<β<β$_c$ interactions with secondary minimum, (c) β$_c$<β<β$_u$ interactions with single equilibrium position at finite distance, (d) β$_u$<β purely repulsive interactions.

The motion of pairs of particles initially in contact in a given field was recorded using an optical microscope with an attached video camera. The experiments where carried out using 50 μm diameter nonmagnetic polystyrene particles produced by Ugelstad's technique [15], in a kerozene-based ferrofluid [16]. The confining cell consisted of two glass plates kept at 70 μm separation by using a few 70 μm diameter spheres as spacers. The cell was placed inside three pairs of coils, and the particle motion was recorded and digitized from the microscopy data. The particles were initially brought into contact by applying a fast oscillating, purely inplane field, and at time $t = 0$ the constant normal field component was added. The particle pair observed was separated from any other particle or spacer by more than 20 diameters, in order to avoid unwanted perturbations. A typical record of the scaled distance as function of the scaled time, obtained for a field $H_\parallel = 14.2$ Oe at $\beta = 0.8$ is shown in Figure 13.

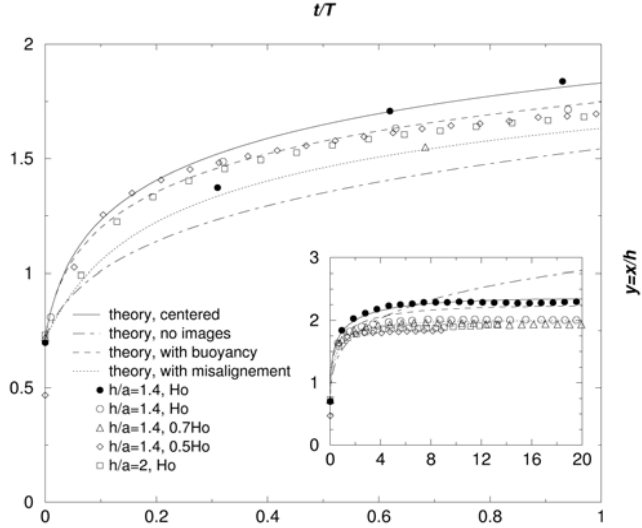

*Figure 13.* Scaled separation *x/h* between a pair of particles as function of the scaled time *t/T* for theory (continuous line) and an experiment obtained with an oscillating inplane field of magnitude $H_\parallel = 14.2$ Oe and a constant normal field $H_\perp = 0.8 H_\parallel$.

In the current case, the time unit is $T = 32$ s, and particles come to the predicted equilibrium distance $r = 2.35 h$ after a few minutes. Comparing this data with the present theory (full line) and with the preexisting expression ignoring the effect of the boundary conditions along the nonmagnetic plates [4,26] (dashed line) is clear in showing the importance of the susceptibility contrast for explaining the existence of the secondary potential minimum.

The dynamical equation ruling the particles in this overdamped regime is obtained by balancing the Stokes drag from the embedding fluid with the magnetic interactions, which leads to $\dot{x} = -u'(x)$, where the dot refers to derivation in with respect to the dimensionless time $t' = t/T$, with $T = 3 T_c \cdot h^5 / a^5$ being the unit time. The function $t'(x)$ was then numerically evaluated as the integral of $-1/u'$ from the initial $a/h$ to the actual $x$ value of the scaled separation, to obtain the full line of Figure 13.

A range of values of the $\beta$ parameter was explored in a series of experiments. Some discrepancy between experiments and theory was found at small particle separation and is believed to be due to the effect of the non point-like character of the magnetic holes, which should generate higher order terms in a multipolar expansion at moderate separations $r/a$. Qualitatively, the dipolar perturbation field of one hole does not fulfill properly the boundary conditions along the surface of another close hole, and a repulsive term corresponding qualitatively to taking into account images of one sphere in the other one, similar to the repulsive effect of images in the plane boundaries on its source particle, becomes sensitive at these short distances.

## 6. Conclusions

In this review we have shown that nonmagnetic particles in magnetized ferrofluids denoted magnetic holes are in many ways ideal model systems to test various forms of particle self assembly and dynamics.

In particular, chaining of magnetic holes show cluster size scaling behavior and for the first time it has been possible to use braid theory and Zipf Relation to describe the dynamics of magnetic holes in ac magnetic fields. Finally, the precise formulation of interactions of magnetic holes in ferrofluid layers has been presented. We have established the effective pair interactions of magnetic holes, submitted to magnetic fields including constant normal components and high frequency oscillating inplane components. Due to the susceptibility contrast along the glass plates, a family of potentials displaying a secondary minimum at finite separation distance has been proven to exist, which should allow trapping of nonmagnetic bodies at tunable distances via the external field.

A system with interactions such as described here, should be relevant for any colloidal suspension of electrically or magnetically polarizable particles constrained in layers. The realization of the detailed effective interaction potentials of this system makes it also a good candidate as an analog model system to study phase transitions, aggregation phenomena in complex fluids, or fracture phenomena in coupled granular/fluid systems.

## Acknowledgements

The Research Council of Norway has in part supported the work described here.